\documentclass[12pt]{article}

\usepackage{amsmath,amssymb,amsfonts,amsthm}
\usepackage{graphicx}
\usepackage{cite}
\usepackage[all]{xy}

\allowdisplaybreaks[3]
\newcommand{\abull}[1]{\noindent
	\parbox[t]{0.04\textwidth}{\hfill $\bullet$\phantom{.}}
	\parbox[t]{0.95\textwidth}{#1 \\[-0.5\baselineskip]} \par}
\newcounter{propositiona}
\newcommand{\propositiona}[1]{\refstepcounter{propositiona}
\textbf{Proposition\, \thepropositiona.}\, {\it #1}}
\newcounter{definitiona}
\newcommand{\definitiona}[1]{\refstepcounter{definitiona}
	\textbf{Definition \thedefinitiona.}\, #1}
\newcounter{remarka}
\newcommand{\remarka}[1]{\refstepcounter{remarka}
	\textbf{Remark \theremarka.}\, #1}
\newcounter{examplea}
\newcommand{\examplea}[1]{\refstepcounter{examplea}
	\textbf{Example \theexamplea.}\, #1}
\begin{document}
	
\hfill {%\small Wednesday 10 August 2022, 13:15}

\large

\thispagestyle{empty}

\begin{center}
{\bf \Large Application of invariants of characteristics to\\[0.3ex] 
construction of solutions without gradient  catastrophe}\\[1.5ex]
{\large \bf  Alexander\! V.\! Aksenov$^{a,b,1}$,\! Konstantin\! P.\! Druzhkov$^{a,c,2}$,\! Oleg\! V.\! Kaptsov$^{d,3}$}\\[1.5ex]
$^a$Lomonosov Moscow State University, 1 Leninskiye Gory,\\
Main Building, 119991 Moscow, Russia\\
$^b$National Research Nuclear University MEPhI,\\
31~Kashirskoe Shosse, 115409 Moscow, Russia\\
$^c$Moscow Institute of Physics and Technology (State
University),\\ 9 Institutskiy per., Dolgoprudny, 141701 Moscow region, Russia\\
$^d$Institute of Computational Modeling, Siberian Branch of RAS,\\ Akademgorodok 50/44, 660036, Krasnoyarsk, Russia\\[0.5ex]
\textit{E-mail: $^1$aksenov@mech.math.msu.su, $^2$Konstantin.Druzhkov@gmail.com, $^3$kaptsov@icm.krasn.ru}
\end{center}

\vspace{1.0ex}

\noindent
\textbf{\Large Abstract}\\[1.0ex]
The one-dimensional system of equations of isentropic gas dynamics is considered. First-order invariants of
characteristics of this system are classified. Second-order invariants of characteristics are classified for polytropic processes. The infinite sequence of Darboux integrable systems is described. The approach to construction of smooth solutions without gradient catastrophe is proposed. Examples of solutions without gradient catastrophe are presented.\\[1.0ex]

\section{Introduction}{\label{Intr}}

It is known that solutions of gas dynamics equations, generally, collapse in finite time, even for smooth initial data.
They can generate so-called gradient catastrophe, i.e. unboundedness of first derivatives. 
In the one-dimensional isentropic case the necessary and sufficient conditions for the gradient
catastrophe can be formulated in terms of initial data~\cite{RozhdYan}. 
However, deriving a concrete solution without gradient catastrophe is a difficult problem. 
So, approaches to construction of solutions without gradient catastrophe are of interest.

The one-dimensional equations of isentropic gas motion admit
Riemann invariants~\cite{Riemann}. For some equations of state, they may also admit additional functions that remain constant on characteristics. Such functions are called invariants of characteristics.
The notion of an invariant of a characteristic is closely related to the Darboux integrability~\cite{Darboux1,Darboux2,Goursat,Vassiliou}.
Let us recall that a system of $m$ first-order differential equations for $m$ unknown functions of two independent variables is Darboux integrable if there are $m$ characteristics, each of which admits at least two functionally independent invariants.

The Darboux integrability of the systems of gas dynamics equations was discussed, for example, in~\cite{Kaptsov1,KaptsZabl,Kaptsov2}. Interpretation of Darboux integrability in terms of integrable systems of hydrodynamic type was found in~\cite{Agafonov}. 
We propose the approach to construction of solutions without gradient catastrophe for the one-dimensional equations of isentropic gas motion in cases of Darboux integrability. 

The paper is organized as follows.
In the section~\ref{Def} we introduce the definition of an invariant of a characteristic and basic nontaions. 
In the section~\ref{ICEul} we classify first-order invariants of characteristics for the barotropic case, as well as second-order invariants for the polytropic case. Important cases $\gamma = 5/3$ and $\gamma = 7/5$ are presented there.
The section~\ref{Sequence} is devoted to the infinite sequence of Darboux integrable systems for different values of $\gamma$.
The main sections~\ref{DerSol} and~\ref{DerSolimp} are devoted to solutions. 
In the section~\ref{DerSol} we consider initial-value problems at entire spatial line, that do not lead to gradient catastrophe.
This section contains the approach to reducing most of these initial-value problems to systems of ordinary differential equations.
It also contains numerical graphs of solutions without gradient catastrophe. The corresponding mass, momentum and energy of gas are finite.
In the section~\ref{DerSolimp} we discuss applications of invariants to
construction of exact solutions. This section contains the example of a solution without gradient catastrophe in implicit analytical form.
The section~\ref{Geom} reveals the geometric nature of the approach. 

\subsection{Basic Equations}\label{BasEq}

In dimensionless variables the one-dimensional equations of isentropic gas motion have the following form
\begin{equation}
\begin{aligned}
&u_t + uu_x + \dfrac{c^2}{\rho}\rho_x = 0\,,\\
&\rho_t + u \rho_x + \rho u_x = 0\,.
\label{Eul_bar}
\end{aligned}
\end{equation}
Here the dimensionless quantities $t$, $x$, $u$, $\rho$ and $c$ are related to the dimensional time $t^*$, spatial coordinate $x^*$, velocity component $u^*$, mass density $\rho^* > 0$ and speed of sound $c^*(\rho^*) > 0$ as follows
$$
t = \dfrac{c_c t^*}{l_c}\,,\qquad x = \dfrac{x^*}{l_c}\,,\qquad u = \dfrac{u^*}{c_c}\,,\qquad \rho = \dfrac{\rho^*}{\rho_c}\,,\qquad
c = \dfrac{c^*}{c_c}\,.
$$
Characteristic length $l_c$ and density $\rho_c$
can be specified in the investigation of a specific problem. The characteristic speed of sound $c_c$ corresponds to the characteristic density.

In case of a polytropic process with $p = \gamma^{-1}\rho^{\gamma}$ the system of equations~\eqref{Eul_bar} takes the form
\begin{equation}
\begin{aligned}
&u_t + uu_x + \rho^{\gamma - 2}\rho_x = 0\,,\\
&\rho_t + u \rho_x + \rho u_x = 0\,.
\label{Eul_pol}
\end{aligned}
\end{equation}
Here the dimensionless quantity $p$ is related to the dimensional pressure $p^*$ by the following formula:
$$
p = \dfrac{p^*}{\rho_c c_c^2}\,.
$$
If $\gamma\neq 1$, mass, momentum and energy conservation laws for the system~\eqref{Eul_pol} can be written as follows.
\begin{align}
\begin{aligned}
&\dfrac{d\,}{dt}\int_{\mathbb{R}}\rho\, dx = -(u\rho)\big|_{-\infty}^{+\infty}\,,\\
&\dfrac{d\,}{dt}\int_{\mathbb{R}}u\rho\, dx = -(u^2\rho + \gamma^{-1}\rho^{\gamma})\big|_{-\infty}^{+\infty}\,,\\
&\dfrac{d\,}{dt}\int_{\mathbb{R}}\Big(\dfrac{u^2\rho}{2} + \dfrac{\rho^{\gamma}}{\gamma(\gamma - 1)}\Big) dx =
-\Big(\dfrac{u^3\rho}{2} + \dfrac{u\rho^{\gamma}}{\gamma - 1}\Big)\Big|_{-\infty}^{+\infty}\,.
\label{Conser}
\end{aligned}
\end{align}

\subsection{Main Results}

Let us formulate the main results of our work:

\abull{Classification of first-order invariants of characteristics for the one-dimensional equations of isentropic gas dynamics is obtained.}

\abull{Classification of second-order invariants of characteristics for the one-dimensional equations of gas dynamics of polytropic processes is obtained.}

\abull{The approach to derivation of solutions without gradient catastrophe for the system of equations~\eqref{Eul_bar} in cases of additional invariants of characteristics is proposed.}

\abull{The systems of equations for polytropic processes with $\gamma = 5/3$ and $\gamma = 7/5$ are reduced to systems of ODEs.
Examples of solutions without gradient catastrophe are presented.}

\abull{Geometric interpretation of the approach is given.}

\section{Basic definitions}\label{Def}

To define invariants of characteristics it is convenient to introduce the operators of total derivatives:
\begin{align*}
&D_x = \dfrac{\partial\ }{\partial x} + u_x\dfrac{\partial\ }{\partial u} + \rho_x\dfrac{\partial\ }{\partial \rho} + u_{xx}\dfrac{\partial\ \ }{\partial u_x} + u_{tx}\dfrac{\partial\ \ }{\partial u_t} + \ldots\,,\\
&D_t = \dfrac{\partial\ }{\partial t} + u_t\dfrac{\partial\ }{\partial u} + \rho_t\dfrac{\partial\ }{\partial \rho} + u_{tx}\dfrac{\partial\ \ }{\partial u_x} + u_{tt}\dfrac{\partial\ \ }{\partial u_t} + \ldots
\end{align*}
Here we assume that these operators are prolonged to derivatives of all orders. Using the system~\eqref{Eul_bar}, we can express $u_t$, $\rho_t$ and all their derivatives. Thus, we obtain the operators of total derivatives due to the system~\eqref{Eul_bar}:
\begin{align*}
&\,\overline{\!D}_x = \dfrac{\partial\ }{\partial x} + u_x\dfrac{\partial\ }{\partial u} + \rho_x\dfrac{\partial\ }{\partial \rho} + u_{xx}\dfrac{\partial\ \ }{\partial u_x} + \rho_{xx}\dfrac{\partial\ \ }{\partial \rho_x} + u_{xxx}\dfrac{\partial\ \ }{\partial u_{xx}} + \ldots\,,\\
&\,\overline{\!D}_t = \dfrac{\partial\ }{\partial t} - \Big(uu_x + \dfrac{c^2}{\rho}\rho_x\Big)\dfrac{\partial\ }{\partial u} - \Big(u \rho_x + \rho u_x\Big)\dfrac{\partial\ }{\partial \rho} - D_x\Big(uu_x + \dfrac{c^2}{\rho}\rho_x\Big)\dfrac{\partial\ \ }{\partial u_x} + \ldots
\end{align*}
The system of equations~\eqref{Eul_bar} admits two characteristics:
$$
dx - (u\pm c ) dt = 0\,.
$$
They can be determined by the derivatives along the corresponding directions
\begin{align}
X_{\pm} = \,\overline{\!D}_t + (u\pm c)\,\overline{\!D}_x\,.
\label{charv}
\end{align}
\!\!\definitiona{A function $f$ (of independent variables, dependent variables and derivatives with respect to $x$ up to some finite order) is an invariant of the characteristic $X_{+}$ if it satisfies the identity}
$$
X_+(f) = 0\,.
$$
Invariants of the characteristic $X_{-}$ are defined similarly. The general definition of an invariant of characteristics of a system of first-order evolution equations with two independent variables is given in~\cite{Kaptsov1}.\\[1.0ex]
\remarka{Any function of invariants of the same characteristic is also invariant of this characteristic.}

\vspace{1.0ex}
Allow us to introduce the notation
$$
\varphi(\rho) = \int_{0}^{\rho}\frac{c(\tilde{\rho})}{\tilde{\rho}}d\tilde{\rho}\qquad (\varphi' > 0)\,.
$$\\[1.0ex]
\examplea{The Riemann invariants~\cite{Riemann}
\begin{equation*}
r = u + \varphi(\rho)\qquad \text{and}\qquad
k = u - \varphi(\rho)
%\label{Riem_inv}
\end{equation*}
are invariants of characteristics for~\eqref{Eul_bar}.}

\section{\label{ICEul}Invariants of characteristics for the one-dimensional gas dynamics}

Let us consider the characteristic vector fields~\eqref{charv} for the system~\eqref{Eul_bar}.
Since their invariants are related by the transformation $t\mapsto -t$, $u\mapsto -u$,
it suffices to classify only invariants of the characteristic field $X_+$.

\subsection{First-order invariants for the system~\eqref{Eul_bar}}

Further we seek first-order $X_+$-invariants, i.e. functions $f(t, x, u, \rho, u_x, \rho_x)$, such that the following identity holds:
$$
X_+(f) = 0\,.
$$

Direct calculation gives us the following system of equations
\begin{align*}
&f_{\rho_x} - \varphi' f_{u_x} = 0\,,\\
&\begin{aligned}
f_t + (u + \rho\varphi')f_x +&\, \rho(u_x - \rho_x \varphi')(\varphi' f_u - f_{\rho}) -{}\\
-&\, (u_x^2 + \rho_x^2\varphi'^2 + 2\rho\rho_x^2\varphi'\varphi'')f_{u_x} - 2u_x\rho_xf_{\rho_x} = 0\,.
\end{aligned}
\end{align*}
This system can be reduced to single classifying equation
\begin{align*}
f_t + rf_x + (\rho\varphi' - \varphi)f_x - \dfrac{2\varphi' + \rho\varphi''}{2\varphi'^{\frac{3}{2}}}q^2f_q = 0\,.
\end{align*}
Here $r = u + \varphi(\rho)$ is the corresponding Riemann invariant for $X_+$;
$q = \rho_x\varphi'^{\frac{3}{2}} + u_x\varphi'^{\frac{1}{2}}$, $f = \tilde{f}(t, x, r, q)$.

It suffices to consider function $\varphi(\rho)$ up to an additive constant.
Also, because of the transformations $x\mapsto \lambda x$, $u\mapsto \lambda u$ in~\eqref{Eul_bar}, 
it suffices to consider function $\varphi(\rho)$ up to a positive multiplicative constant.
Besides, the Riemann invariant $r = u + \varphi(\rho)$ exists for any function $\varphi(\rho)$, so we are only interested in cases where additional invariants of characteristics exist. 

The classifying equation has a special structure. Namely, its left-hand side is a sum of the form 
$$
\sum_i w_i(\rho)\cdot v_i(t, x, r, q)\,.
$$
Using this special structure, we obtain the following five cases.

\noindent
\textbf{\mathversion{bold}Case 1. $\varphi = \rho$.}
In this case we have
$$
p = \dfrac{\rho^3}{3}\,.
$$
The additional invariants are
\begin{align*}
I_1 = x - (u + \rho)t\,,\qquad I_2 = \dfrac{1}{u_x + \rho_x} - t\,.
\end{align*}
\textbf{\mathversion{bold}Case 2. $\varphi = -(\rho + C)^{-1}$.}
In this case we have
$$
p = -\dfrac{3\rho^2 + 3C\rho + C^2}{3(\rho + C)^3}\,.
$$
The additional invariant is
\begin{align*}
I = Ct - \dfrac{(\rho + C)^3}{\rho_x + (\rho + C)^2u_x}\,.
\end{align*}
\textbf{\mathversion{bold}Case 3. $\varphi = 3\cdot\sqrt[3]{\rho + C}$, where $\rho(x, t) + C \neq 0$.}
In this case we have
$$
p = \dfrac{3}{5}\cdot\dfrac{\rho^2 - 3C\rho - 9C^2}{\sqrt[3]{\rho + C}}\,.
$$
The additional invariant is
\begin{align*}
I = x - (u + 3\cdot\sqrt[3]{\rho + C})t + \dfrac{3(\rho + C)}{\rho_x + \sqrt[3]{(\rho + C)^2}\,u_x}\,.
\end{align*}
\textbf{\mathversion{bold}Case 4. $\varphi(\rho)$ is implicitly defined by the equation
$$
C_1\rho(e^{-\varphi} + C_2 e^{\varphi}) =
(\varphi + 1)e^{-\varphi} + C_2(\varphi - 1)e^{\varphi}\,,
$$
where $C_1 > 0$, $C_2\neq 0$ and $\varphi(\rho(x, t)) \neq C_1\rho(x, t)$.}\\
In this case the additional invariant is
\begin{align*}
I = x - (u + \varphi)t + \dfrac{(\varphi - C_1\rho)^3}{C_1\rho_x + (\varphi - C_1\rho)^2u_x}\,.
\end{align*}
\textbf{\mathversion{bold}Case 5. $\varphi(\rho)$ is implicitly defined by the equation
$$
C_1\rho(\cos\varphi - C_2\sin\varphi) = (C_2\varphi + 1)\sin\varphi + (C_2 - \varphi)\cos\varphi\,,
$$
where $C_1 > 0$.}\\
In this case the additional invariant is
\begin{align*}
I = x - (u + \varphi)t +
\dfrac{(\varphi + C_1\rho)^3}{C_1\rho_x + (\varphi + C_1\rho)^2u_x}\,.
\end{align*}

\subsection{Second-order invariants for the system~\eqref{Eul_pol}}

Now we consider the system of equations~\eqref{Eul_pol} and look for additional cases of existence of second-order $X$-invariants. Direct calculations lead us to the following classifying equation
\begin{align*}
64(\gamma - 1)(f_t + r f_x) + 64(\gamma - 3)\rho^{\frac{\gamma - 1}{2}}f_x &+ 16(1 - \gamma^2)\rho^{\frac{3 - \gamma}{4}}(a^2f_{a} + 3ab f_{b}) +{}\\
&+ (3\gamma^3 - 7\gamma^2 - 7\gamma + 3)\rho^{\frac{5 - 3\gamma}{4}}a^3f_{b} = 0\,.
\end{align*}
Here $\gamma\neq 1$, $f = \tilde{f}(t, x, r, a, b)$,
\begin{align*}
&a = \rho^{\frac{\gamma - 3}{4}}u_x + \rho^{\frac{3(\gamma - 3)}{4}}\rho_x\,,\qquad b = \rho^{\frac{\gamma - 3}{4}}a_x + \dfrac{\gamma^2 - 2\gamma - 3}{16(\gamma - 1)\rho}\big(u_x + \rho^{\frac{\gamma - 3}{2}}\rho_x\big)^2.
\end{align*}
Then the following two (additional) cases arise here:

\noindent
\textbf{\mathversion{bold}Case 6. $\gamma = 1/3$.}\\
In this case the additional invariant is
\begin{align*}
I = \dfrac{b}{a^3}\,.
\end{align*}
\textbf{\mathversion{bold}Case 7. $\gamma = 7/5$.}\\
In this case the additional invariant is
\begin{align*}
I = x - (u + 5\rho^{\frac{1}{5}})t - \dfrac{25b}{3a^3}\,.
\end{align*}\\[1.0ex]
\remarka{In the cases $1$--$5$ there are additional second-order invariants. They have the form $r_x^{-1}\,\overline{\!D}_x(I)$.
Generally speaking, the operator $r_x^{-1}\,\overline{\!D}_x$ acts on the set of $X_+$-invariants. This fact follows from the identity
$$
X_+ \Big(\dfrac{1}{r_x}\,\overline{\!D}_x(\psi)\Big) = \dfrac{1}{r_x}\,\overline{\!D}_x(X_+(\psi))\,,
$$
which is valid for any function $\psi$.}

\section{On the infinite sequence of the Darboux integrable systems}\label{Sequence}
Since Riemann's classical work~\cite{Riemann} it is known that the system of equations~\eqref{Eul_pol} is related to the Euler-Poisson-Darboux equation.
Let us put
$$
\alpha = \dfrac{\gamma + 1}{\gamma - 1}\qquad\qquad (\gamma \neq 1\,,\ \alpha\neq 1)
$$
(for simplicity of formulas)
and consider the system~\eqref{Eul_pol} in terms of $\alpha$. Further we denote it by GDE($\alpha$). This system is related to the Euler-Poisson-Darboux equation
\begin{align*}
\text{EPD}(\alpha)\colon\qquad v^{(\alpha)}_{yy} + \dfrac{\alpha}{y}v^{(\alpha)}_y - v^{(\alpha)}_{\tau\tau} = 0
\end{align*}
by the mapping (differential covering~\cite{VinKr}) $s^{(\alpha)}\colon \,\text{GDE}(\alpha) \to \,\text{EPD}(\alpha)$,
\begin{align*}
s^{(\alpha)}\colon\quad
\begin{aligned}
&\tau = \dfrac{u}{\alpha - 1}\,,\qquad y = \rho^{\frac{1}{\alpha - 1}}\,,\qquad v^{(\alpha)} = t\,,
\qquad\ldots
\end{aligned}
%\label{subst}
\end{align*}
In addition, there are four types of first order linear mappings between the Euler-Poisson-Darboux equations~\cite{Aksenov1}. Schematically, all this can be depicted in the diagram.
\begin{align*}
\xymatrix
{
\mathrm{EPD}\, (\alpha - 2) \ar[rd]^{\sigma_2} &\ \mathrm{GDE}\, (\alpha) \ar[d]^{s^{(\alpha)}} &\ \mathrm{EPD}\, (\alpha) \ar[ld]_{\sigma_3}\\
\mathrm{EPD}\, (2 - \alpha) \ar[r]^{\sigma_1} & \mathrm{EPD}\, (\alpha) & \mathrm{EPD}\, (\alpha + 2) \ar[l]_{\sigma_4}\\
}
\end{align*}
For example, the mappings determined by the formulas
\begin{align*}
&s_1^{(\alpha)}\colon\ \, v^{(\alpha)} = y^{1 - \alpha} v^{(2 - \alpha)}\,,\qquad
&&s_2^{(\alpha)}\colon\ \, v^{(\alpha)} = y^{-1}v^{(\alpha - 2)}_y\,,\\
&s_3^{(\alpha)}\colon\ \, v^{(\alpha)} = \tilde{v}^{(\alpha)}_{\tau}\,,  
&&s_4^{(\alpha)}\colon\ \, v^{(\alpha)} = (\alpha + 1)v^{(\alpha + 2)} + yv^{(\alpha + 2)}_y
\end{align*}
are of the types $\sigma_1$, $\sigma_2$, $\sigma_3$ and $\sigma_4$ respectively
(these mappings preserve the independent variables).

It is easy to see that the mappings $s^{(\alpha)}$, $s^{(\alpha)}_1$, $s^{(\alpha)}_2$, $s^{(\alpha)}_3$, $s^{(\alpha)}_4$ transform
characteristics to characteristics (via pull-backs). For instance, consider the characteristic $d\tau + dy = 0$ of EPD($\alpha$). Using the expressions
$$
\tau = \dfrac{u(t, x)}{\alpha - 1}\,,\qquad y = \rho(t, x)^{\frac{1}{\alpha - 1}}
$$
from $s^{(\alpha)}$, we obtain the following relation:
\begin{align*}
\dfrac{u_x + \rho^{\frac{2 - \alpha}{1 - \alpha}}\rho_x}{\alpha - 1}\,\big(dx - (u + \rho^{\frac{1}{\alpha - 1}})dt\big) = 0\,.
\end{align*}
Therefore, the mapping $s^{(\alpha)}$ transforms characteristics of EPD($\alpha$) to characteristics of~\eqref{Eul_pol}. 

Hereby, we obtain the following\\[1.0ex]
\propositiona{The mappings $\sigma_i$ and $s^{(\alpha)}$ transform invariants of characteristics to invariants of characteristics.}\\[1.0ex]
\remarka{The mapping $s^{(\alpha)}$ transforms invariants into invariants independent of $x$.}\\[1.0ex]
So, we have the infinite sequence of the Darboux integrable systems of the form~\eqref{Eul_pol} for different values of $\gamma$.

\section{Derivation of solutions}\label{DerSol}

%We propose three approaches to derivation of smooth solutions without gradient catastrophe for the system of equations~\eqref{Eul_bar} in cases of additional invariants of characteristics. These approaches are demonstrated in various cases of existence of additional invariants of the system~\eqref{Eul_pol}.
It is handly to rewrite the system~\eqref{Eul_pol} in terms of the Riemann invariants (for $\gamma \neq 1$):
\begin{align}
\begin{aligned}
&r_t + \dfrac{(1 + \gamma)r + (3 - \gamma)k}{4}\, r_x = 0\,,\\
&k_t + \dfrac{(1 + \gamma)k + (3 - \gamma)r}{4}\, k_x = 0\,.
\end{aligned}
\label{Riem}
\end{align}\\[1.0ex]
\remarka{\label{vice}
The transformation $r\mapsto k$, $k\mapsto r$ maps $X_+$-invariants of~\eqref{Riem} to $Y$-invatiants and vice versa.}\\[1.0ex]
Further we examine a Cauchy problem with a smooth initial data at entire spatial line $x\in\mathbb{R}$ for the system~\eqref{Riem}:
\begin{align}
r|_{t=0} = r_0(x)\,,\qquad\qquad k|_{t=0} = k_0(x)\,.
\label{Init}
\end{align}
It is known (see~\cite{RozhdYan}) that for $-1 < \gamma \leqslant 3$ the necessary and sufficient conditions for the impossibility of 
gradient catastrophe in~\eqref{Riem},~\eqref{Init} are
\begin{equation}
r_0'(x) \geqslant 0\,,\qquad k_0'(x) \geqslant 0\,.
\label{Init_good}
\end{equation}

Application of invariants to construction of solutions without gradient catastrophe for~\eqref{Eul_bar} can be formulated as follows.
Suppose there is an additional $X_+$-invariant $I_+$. Let us add an equation of the form $I_+ = g(r)$ to the system~\eqref{Eul_bar}. One can consider the characteristic field $X_-$ due to this overdetermined system. Computation of the flow of such a vector field amounts to solving a finite closed system of ODEs. This flow allows to derive solutions to initial-value problems.\\[1.0ex]
\remarka{We can use smooth initial conditions~\eqref{Init} satisfying~\eqref{Init_good} to define function $g$. 
At least if $r_0(x)$ is an injection and $I_+(0, x, r_0(x), k_0(x), \ldots)$ is defined for all $x\in\mathbb{R}$, one can find the inverse to $r_0(x)$ and define $g$. The degree of smoothness required for the initial conditions is determined by the invariant at hand.}\\[1.0ex]
In the following, we demonstrate this approach with a few examples.

\subsection{Case $p = \rho^3/3$}

In this case the system of equations~\eqref{Riem} consists of two Hopf equations
\begin{align*}
&r_t + rr_x = 0\,,\\
&k_t + kk_x = 0\,.
\end{align*}
Then smooth solutions for $p = \rho^3/3$ can be easily constructed
without using invariants of characteristics. However, it seems to us
instructive to demonstrate the approach in this case.

Consider invariant $I_1 = x - rt$ and the relation $x - rt = g(r)$.
This relation is simply a general solution to the Hopf equation. 
Suppose the variable $r$ can be eliminated from the equation $x - rt = g(r)$. Then the flow of the characteristic field $X_-$ due to the system of equations
\begin{align*}
&r_t + rr_x = 0\,,\\
&k_t + kk_x = 0\,,\\
&x - rt = g(r)
\end{align*}
is determined by the following ODE system
\begin{align*}
&\dot{x} = k\,,\qquad \dot{k} = 0\,.
\end{align*}
It transforms the initial curve as follows
\begin{align*}
&x' = x + k_0(x)t\,,\\
&k' = k_0(x)\,.
\end{align*}
The last steps are to rewrite the solution $k' = k_0(x)$ in terms of $t$, $x'$ and omit the primes.
So, here we just solved two Hopf equations (for $r$ and $k$) in two different ways.

%Below we prefer to deal with $Y$-invariants (it is convenient from a technical point of view). 

\subsection{Case $p = \rho^\frac{5}{3}\cdot 3/5$}
In this case the additional $X_+$-invariant is 
\begin{equation*}
I_+ = x - rt + \dfrac{r - k}{2r_x}\,.
\end{equation*}
Then the additional $X_-$-invariant is
\begin{equation*}
I_- = x - kt + \dfrac{k - r}{2k_x}\,.
\end{equation*}
Below we prefer to deal with $X_-$-invariants (it is more convenient from a technical point of view).

Let us consider the overdetermined system of equations
\begin{align}
\begin{aligned}
&r_t + \dfrac{2r + k}{3}\, r_x = 0\,,\\
&k_t + \dfrac{2k + r}{3}\, k_x = 0\,,\\
&x - kt + \dfrac{k - r}{2k_x} = h(k),
\end{aligned}
\label{G53}
\end{align}
where $h=h(k)$ is an arbitrary function.\\[1.0ex]
\remarka{The system under consideration reduces to the first-order scalar equation
$$
k_t + kk_x + \dfrac{2}{3}\big(x - kt - h(k)\big)k_x^2 = 0\,.
$$
}\\[-1.0ex]
We can express the variable $k_x$ from the last equation of~\eqref{G53}: 
$$
k_x = \dfrac{r - k}{2\big(x - kt - h(k)\big)}\,.
$$
The compatibility condition $k_{tx} = k_{xt}$ does not lead to nontrivial constraints. 
Then the flow of the characteristic field $X_+$ due to the system of equations~\eqref{G53}
is determined by the following ODE system
\begin{align}
\begin{aligned}
&\dot{x} = \dfrac{2r + k}{3}\,,\qquad \dot{r} = 0\,,\qquad \dot{k} = \dfrac{(r - k)^2}{6\big(x - kt - h(k)\big)}\,.
\end{aligned}
\label{restr}
\end{align}

As examples, consider the initial conditions
\begin{align}
\begin{aligned}
&(a)\colon\quad r_0(x) = x + \dfrac{1}{1 + x^2}\,,\qquad k_0(x) = x\,,\\
&(b)\colon\quad r_0(x) = \arctan(x) + \dfrac{1}{1 + x^2}\,,\qquad k_0(x) = \arctan(x)\,.
\end{aligned}
\label{initsol}
\end{align}
Here the initial velocity component and mass density are
\begin{align*}
&(a)\colon\quad u(0, x) = x + \dfrac{1}{2(1 + x^2)}\,,\qquad \rho(0, x) = \dfrac{1}{6^3\cdot (1 + x^2)^3}\,,\\
&(b)\colon\quad u(0, x) = \arctan(x) + \dfrac{1}{2(1 + x^2)}\,,\qquad \rho(0, x) = \dfrac{1}{6^3\cdot (1 + x^2)^3}\,.
\end{align*}
\textit{These initial conditions are chosen so that the mass, momentum and energy from~\eqref{Conser} are finite at $t = 0$.
The initial speed in (b) is bounded on $\mathbb{R}$.}
The constraint 
$$
I_-(0, x, r_0(x), k_0(x), \ldots) = h(k_0(x))
$$ 
allows us to find $h(k)$:
\begin{align*}
&(a)\colon\quad h(k) = k - \dfrac{1}{2(1 + k^2)}\,,\qquad\qquad (b)\colon\quad h(k) = \tan(k) - \dfrac{1}{2}\,.
\end{align*}\\[1.0ex]
\remarka{\label{badcond}
There are initial conditions for which it is impossible to choose the corresponding function $h$. For example: if $k_0'$ vanishes at some point (or on some 
interval) where $k_0(x)$ differs from $r_0(x)$, there is no suitable function $h$.
However, in this case one can apply $X_+$-invariants in a similar way (if possible).}\\[1.0ex]
The results of the numerical investigation for~\eqref{restr} in the cases (a) and (b) are presented
in the figures~\ref{fig1},~\ref{fig2}.
\begin{figure}[h!]  
\includegraphics[width=0.5\textwidth]{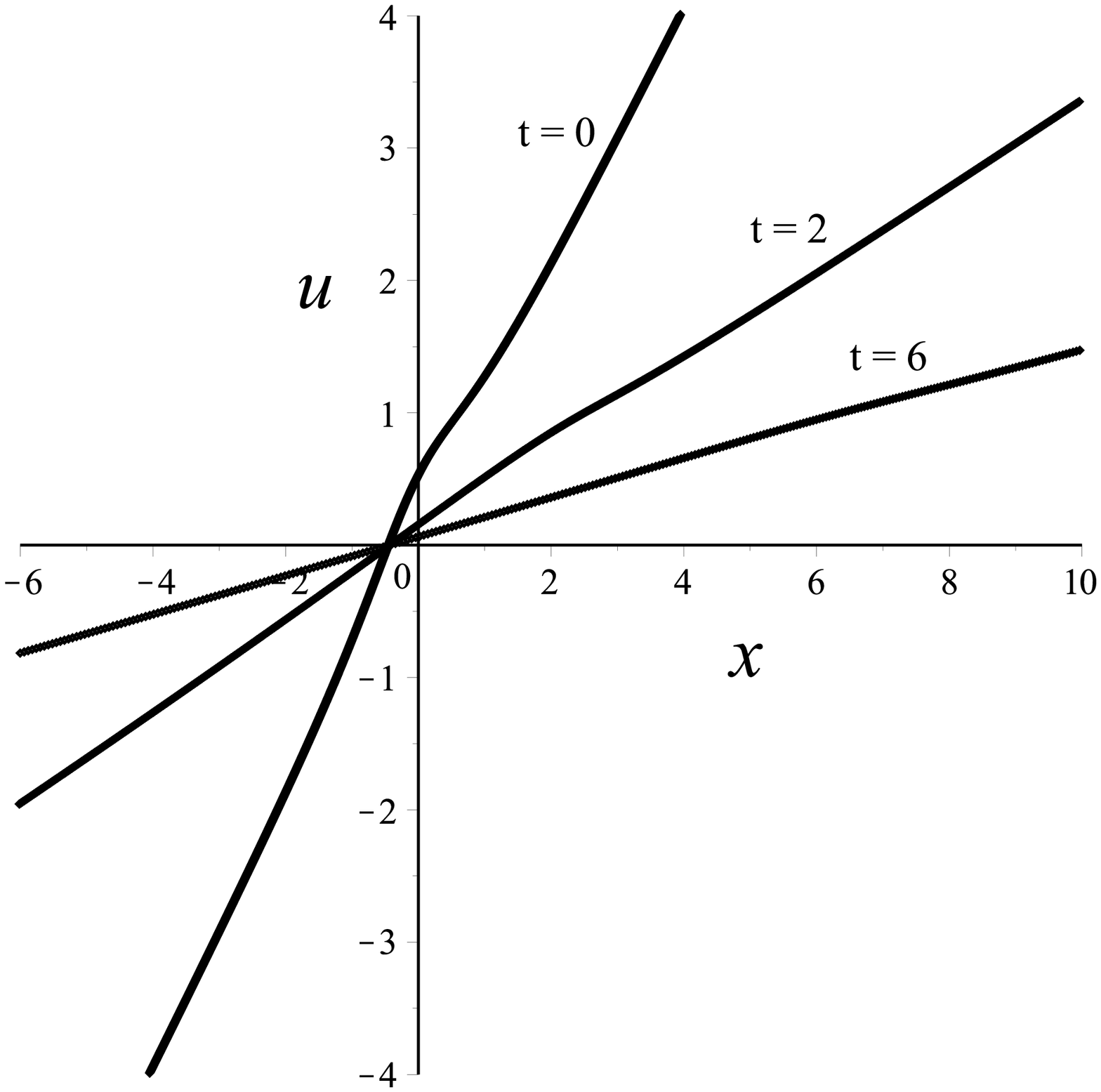}
\hfill
\raisebox{9pt}{\includegraphics[width=0.5\textwidth]{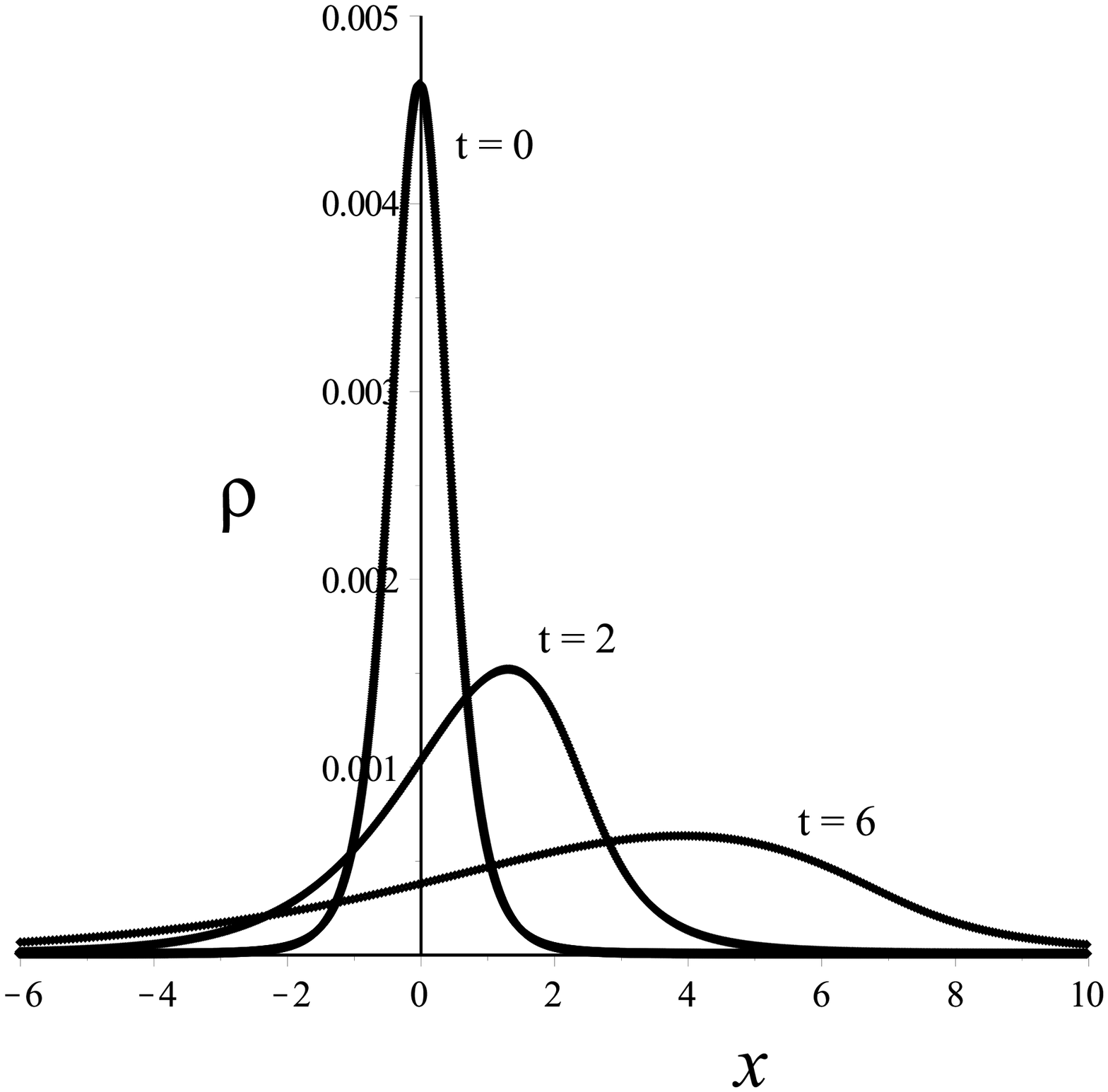}}\ \
\caption{The problem (a) for $\gamma = 5/3$ at $t = 0, 2, 6$.}
\label{fig1}
\end{figure}

Figure 1: The initial mass density is symmetrical about the vertical axis.
Further the density becomes asymmetric about its maximum. It slopes to the right. However, 
the density spread along the $x$-axis faster than it slopes to the right.
\begin{figure}[h!]  
\includegraphics[width=0.5\textwidth]{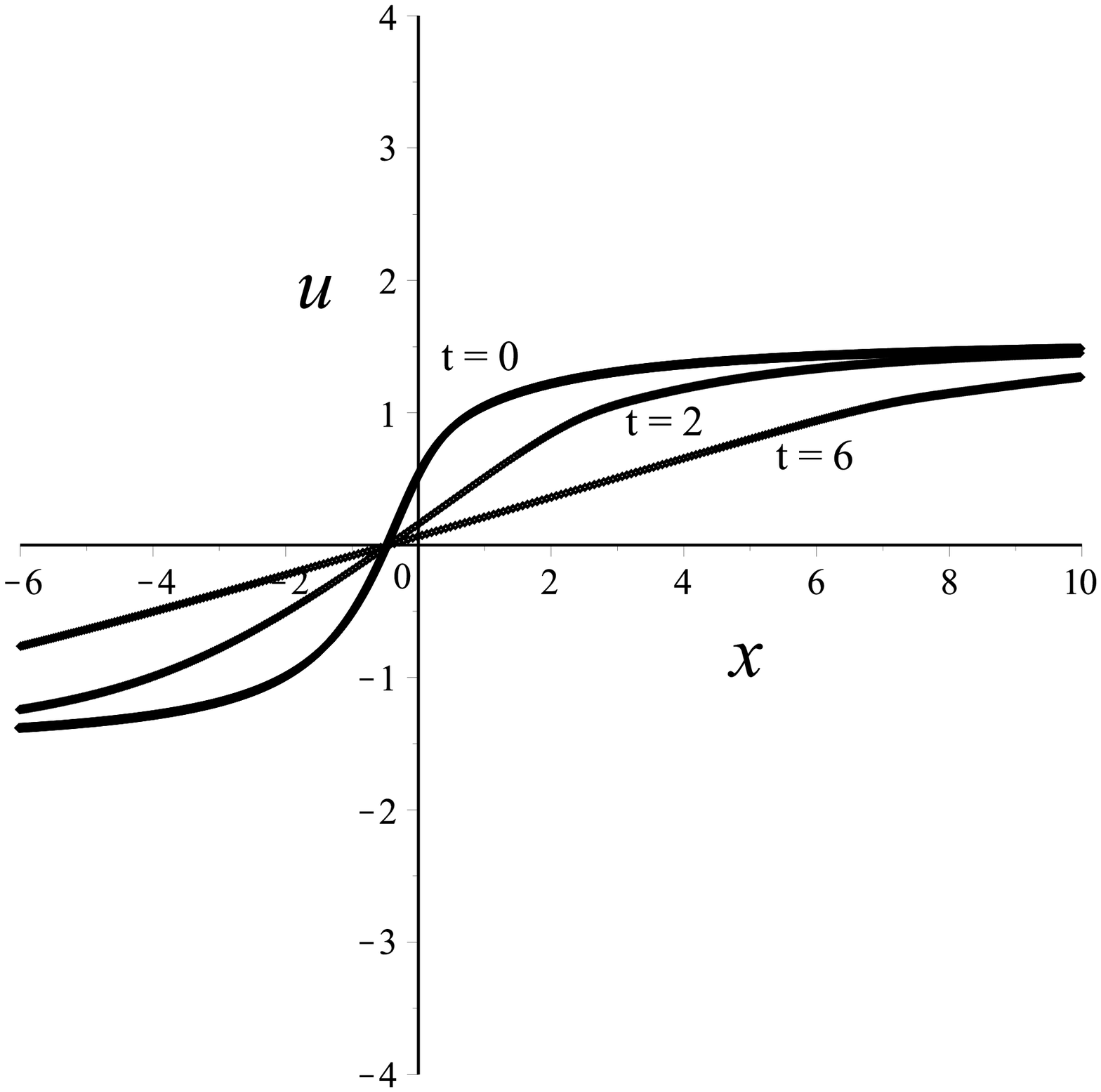}
\hfill
\raisebox{9pt}{\includegraphics[width=0.5\textwidth]{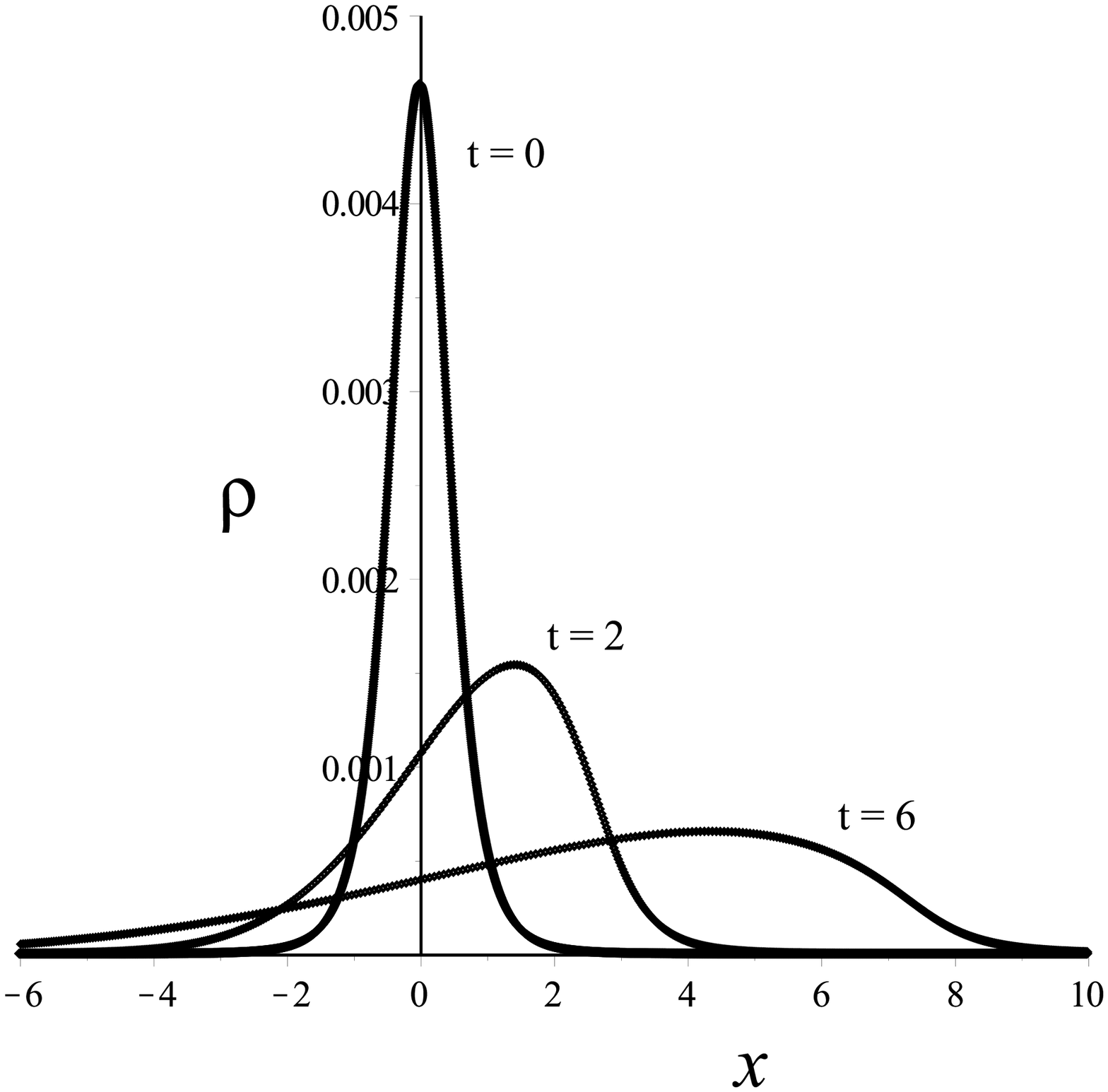}}\ \
\caption{The problem (b) for $\gamma = 5/3$ at $t = 0, 2, 6$.}
\label{fig2}
\end{figure}

Figure 2: The initial velocity in (b) slightly differs from the initial velocity in (a) near to the initial density maximum $x = 0$. As a result, the difference between the mass densities in (a) and (b) is not essential for $t > 0$ (near to the maximums).

\subsection{Case $p = \rho^\frac{7}{5}\cdot 5/7$}
In this case the additional $X_+$-invariant is 
\begin{equation*}
I_+ = x - rt + (r - k)\dfrac{2r_x(4r_x - k_x) - (r - k)r_{xx}}{12 r_x^3}\,.
%\label{InvI}
\end{equation*}
Then the additional $X_-$-invariant is
\begin{equation*}
I_- = x - kt + (k - r)\dfrac{2k_x(4k_x - r_x) - (k - r)k_{xx}}{12 k_x^3}
%\label{InvJ}
\end{equation*}
and we deal with the overdetermined system of equations
\begin{align}
\begin{aligned}
&r_t + \dfrac{3r + 2k}{5}\, r_x = 0\,,\\
&k_t + \dfrac{3k + 2r}{5}\, k_x = 0\,,\\
&x - kt + (k - r)\dfrac{2k_x(4k_x - r_x) - (k - r)k_{xx}}{12 k_x^3} = h(k).
\end{aligned}
\label{G75}
\end{align}
One can express the variable $r_x$ from the last equation: 
$$
r_x = 4k_x + \dfrac{6\big(x - kt - h(k)\big)k_x^2}{k - r} - \dfrac{(k - r)k_{xx}}{2k_x}\,.
$$
Now the system~\eqref{G75} amounts to the system of equations
\begin{align*}
&r_x = 4k_x + \dfrac{6\big(x - kt - h(k)\big)k_x^2}{k - r} - \dfrac{(k - r)k_{xx}}{2k_x}\,,\\
&r_t = -\dfrac{3r + 2k}{5}\Big(4k_x + \dfrac{6\big(x - kt - h(k)\big)k_x^2}{k - r} - \dfrac{(k - r)k_{xx}}{2k_x}\Big)\,,\\
&k_t = - \dfrac{3k + 2r}{5}\, k_x\,.
\end{align*}
Let us stress that \textit{this system of equations does not amount to a scalar first-order equation for either $r$ or $k$.}

The compatibility condition $r_{tx} = r_{xt}$ does not lead to nontrivial constraints. 
Then the flow of the characteristic field $X_+$ due to the system of equations~\eqref{G75}
is determined by the following ODE system
\begin{align}
\begin{aligned}
&\dot{x} = \dfrac{3r + 2k}{5}\,,\qquad \dot{r} = 0\,,\qquad \dot{k} = - \dfrac{k - r}{5} k_x\,,\\
&\dot{k}_x = - \Big(\dfrac{12k_x^3\big(x - kt - h(k)\big)}{5(k - r)} + \dfrac{11}{5}k_x^2\Big)\,.
\end{aligned}
\label{Xfield}
\end{align}\\[1.0ex]
\remarka{Since we considering the second-order invariant, 
we need initial conditions of at least $C^2$ class.}

\vspace{1.0ex}
Let us use the initial conditions~\eqref{initsol} again.
The corresponding initial velocity component and mass density become
\begin{align*}
&(a)\colon\quad u(0, x) = x + \dfrac{1}{2(1 + x^2)}\,,\qquad \rho(0, x) = \dfrac{1}{10^5\cdot (1 + x^2)^5}\,,\\
&(b)\colon\quad u(0, x) = \arctan(x) + \dfrac{1}{2(1 + x^2)}\,,\qquad \rho(0, x) = \dfrac{1}{10^5\cdot (1 + x^2)^5}\,.
\end{align*}
It is easy to verify that the mass, momentum and energy from~\eqref{Conser} are finite at $t = 0$.
The constraint 
$$
I_-(0, x, r_0(x), k_0(x), \ldots) = h(k_0(x))
$$ 
allows to find $h(k)$:
\begin{align*}
&(a)\colon\quad h(k) = \dfrac{6k^7 + 18k^5 - 3k^4 + 18k^3 - 6k^2 + 4k - 3}{6(1 + k^2)^3}\,,\\
&(b)\colon\quad h(k) = \dfrac{6\tan(k)^3-3\tan(k)^2+5\tan(k)-3}{6(1 + \tan(k)^2)}\,.
\end{align*}
Here we numerically calculate the solutions of~\eqref{Xfield} for (a) and (b). The results are presented
in the figures~\ref{fig3},~\ref{fig4}.
\begin{figure}[h!]  
\includegraphics[width=0.5\textwidth]{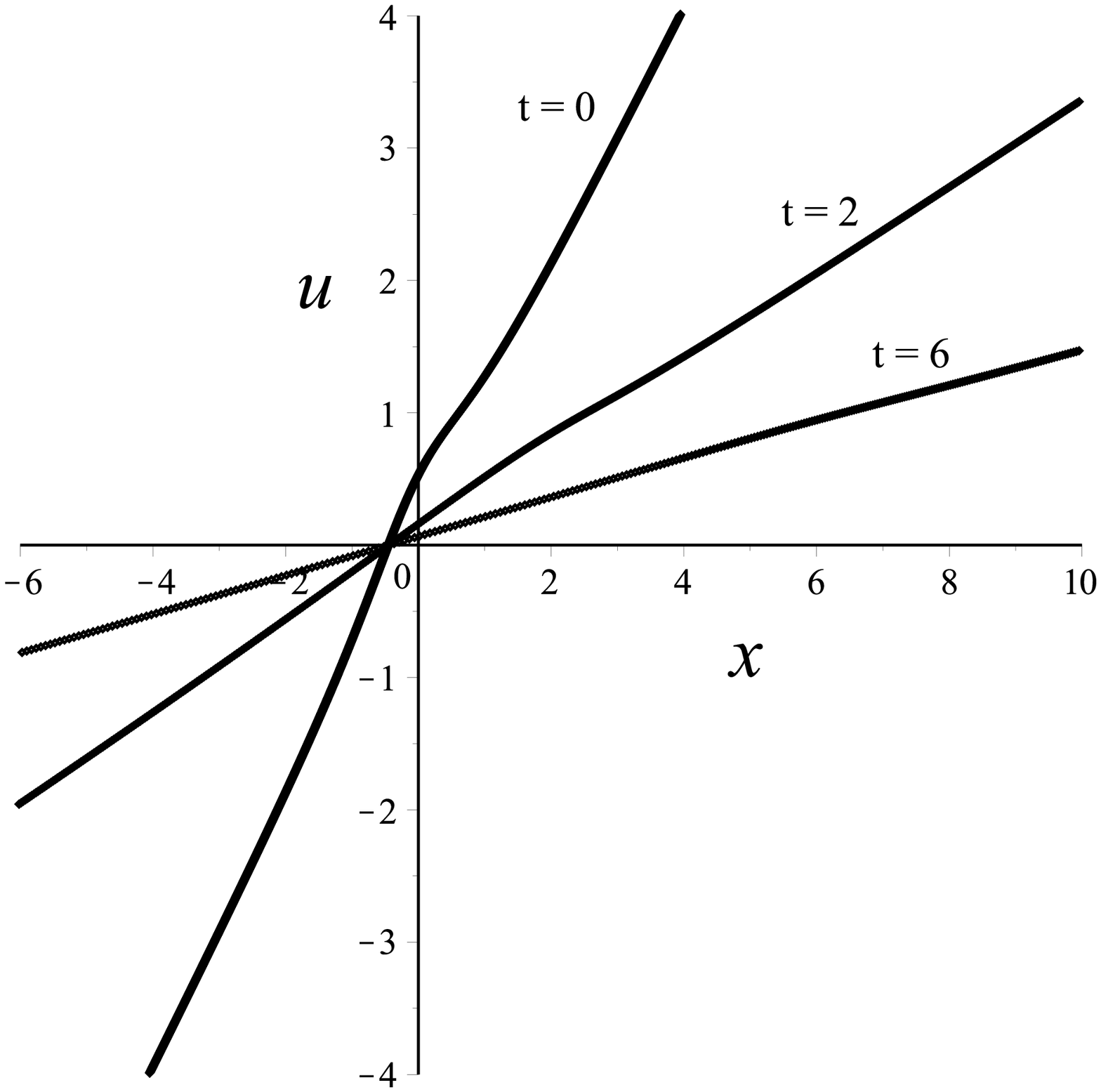}
\hfill
\raisebox{9pt}{\includegraphics[width=0.5\textwidth]{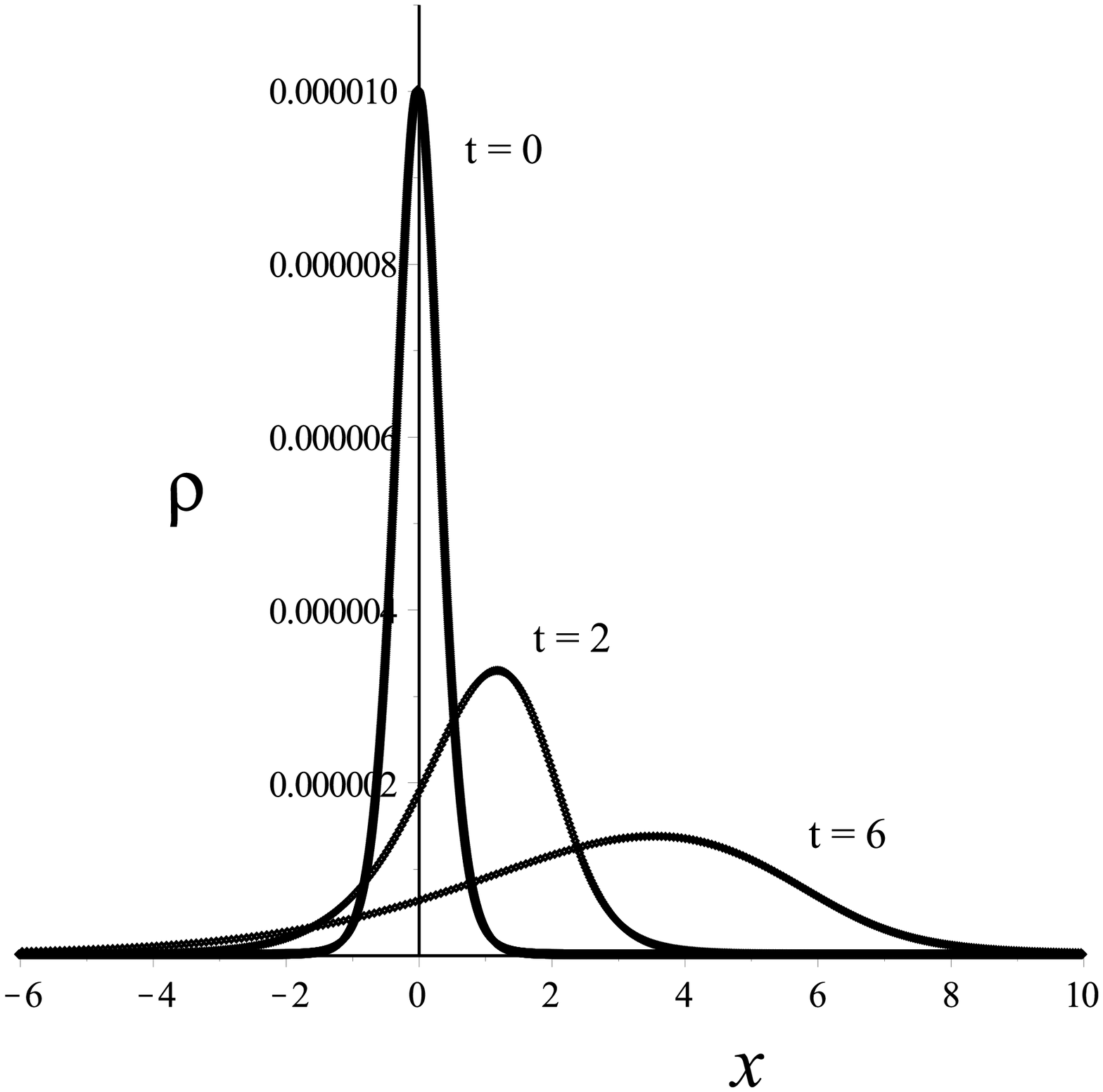}}\ \
\caption{The problem (a) for $\gamma = 7/5$ at $t = 0, 2, 6$.}
\label{fig3}
\end{figure}

Figure 3: The initial mass density is symmetrical about the vertical axis.
Further the density becomes asymmetric about its maximum. It slopes to the right. However, 
the density spread along the $x$-axis faster than it slopes to the right.
\begin{figure}[h!]  
\includegraphics[width=0.5\textwidth]{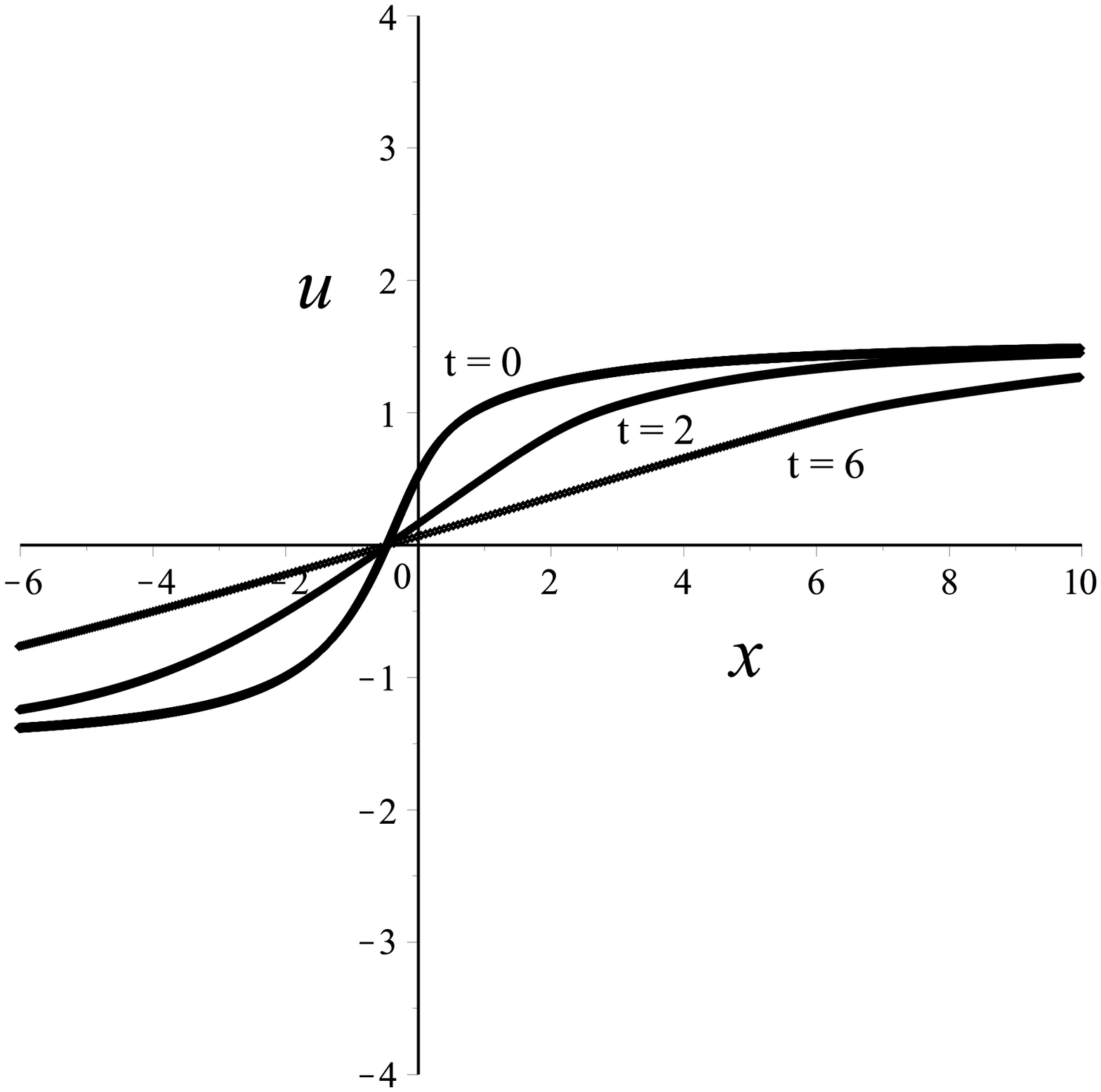}
\hfill
\raisebox{9pt}{\includegraphics[width=0.5\textwidth]{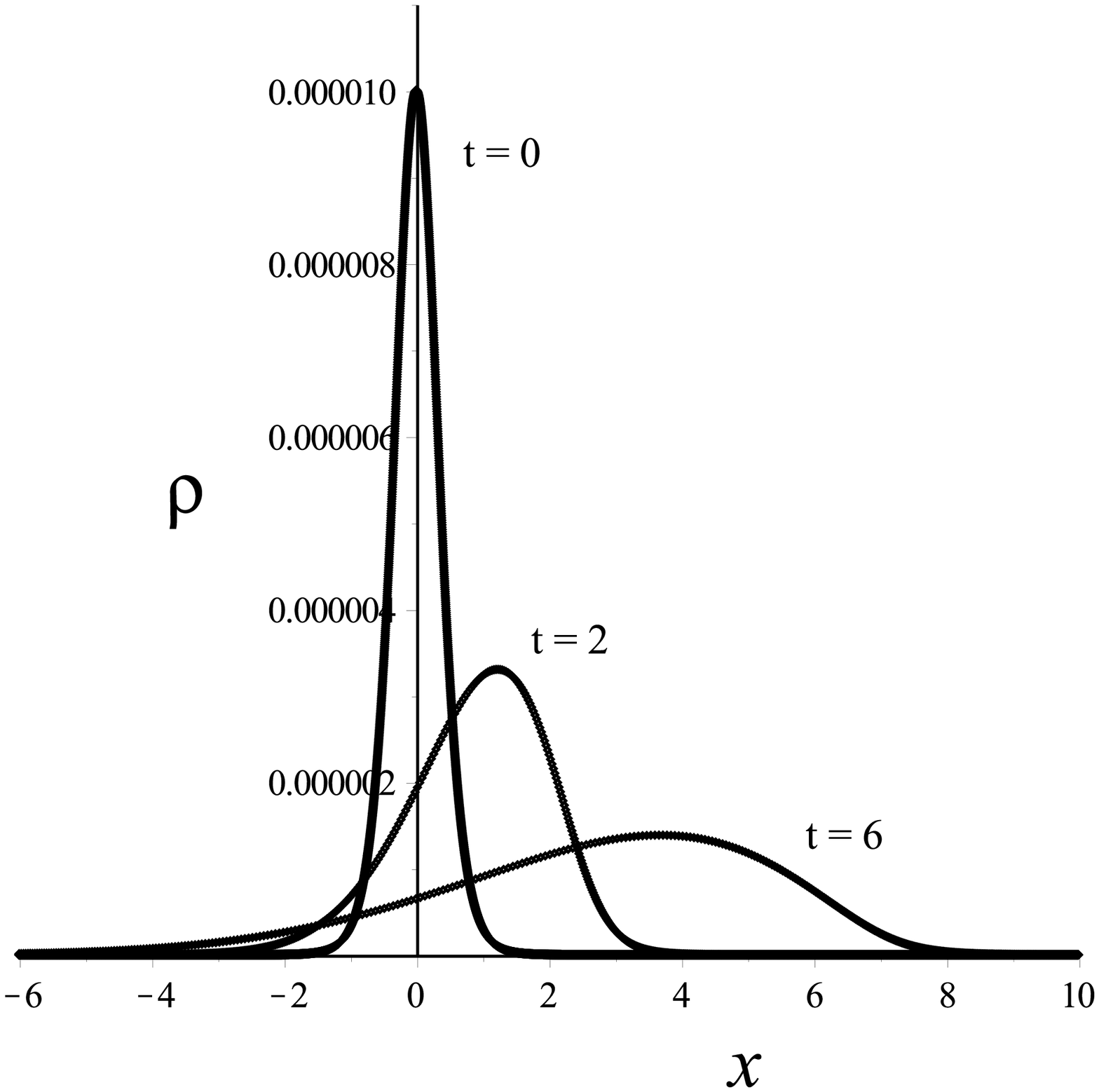}}\ \
\caption{The problem (b) for $\gamma = 7/5$ at $t = 0, 2, 6$.}
\label{fig4}
\end{figure}

Figure 4: The initial velocity in (b) slightly differs from the initial velocity in (a) near to the initial density maximum $x = 0$. As a result, the difference between the mass densities in (a) and (b) is not essential for $t > 0$ (near to the maximums).

\section{On other approaches to derivation of solutions}\label{DerSolimp}

\subsection{Derivation of exact solutions in explicit form}
%The first approach is based on using invariants of both characteristic distributions. It allows us to reduce some initial value problems for~\eqref{Eul_bar} to systems of ordinary differential equations. Such ODE systems are related to explicit forms of desired solutions.
Let us consider the system of equations~\eqref{Eul_bar} and suppose there are additional $X_+$-invariant $I_+$ and $X_-$-invariant $I_-$. We can add equations of the form $I_+ = g(r)$ and $I_- = h(k)$ to the system~\eqref{Eul_bar}. Then the corresponding overdetermined system can be regarded as the ODE system with parameter $x$. It allows to derive solutions of~\eqref{Eul_bar} in explicit form.\\[1.0ex]
\examplea{Consider the system~\eqref{Riem} with $\gamma = 1/3$. In this case the additional invariants are
$$
I_+ = \dfrac{r_{xx}(k - r) + 2r_xk_x}{(k - r)^3r_x^3}\,,\qquad I_- = \dfrac{k_{xx}(r - k) + 2k_xr_x}{(r - k)^3k_x^3}\,,
$$
hence we obtain the following overdetermined system of equations
\begin{align*}
\begin{aligned}
&r_t + \dfrac{r + 2k}{3}\, r_x = 0\,,\\
&k_t + \dfrac{k + 2r}{3}\, k_x = 0\,,
\end{aligned}\qquad\qquad
\begin{aligned}
&\dfrac{r_{xx}(k - r) + 2r_xk_x}{(k - r)^3r_x^3} = g(r)\,,\\
&\dfrac{k_{xx}(r - k) + 2k_xr_x}{(r - k)^3k_x^3} = h(k)\,.
\end{aligned}
%\label{G13o}
\end{align*}
We can express the variables $r_t$, $k_t$, $r_{xx}$, $k_{xx}$ and their derivatives from this system. Therefore, we obtain the following ODE system:
\begin{align}
\begin{aligned}
&\dot{r} = -\dfrac{r + 2k}{3}\, r_x,\\
&\dot{k} = -\dfrac{k + 2r}{3}\, k_x,
\end{aligned}
\begin{aligned}
&\dot{r}_x = -\dfrac{r_x + 2k_x}{3}\, r_x - \dfrac{r + 2k}{3}\Big(\!(k - r)^2r_x^3\, g(r) - \dfrac{2r_xk_x}{k - r}\!\Big),\\
&\dot{k}_x = -\dfrac{k_x + 2r_x}{3}\, k_x - \dfrac{k + 2r}{3}\Big(\!(r - k)^2k_x^3\, h(k) - \dfrac{2k_xr_x}{r - k}\!\Big).
\end{aligned}
\label{ODE}
\end{align}
}\\[1.0ex]
\propositiona{Any one-parameter family of explicit solutions to~\eqref{ODE}
(with parameter $x$) determine explicit solution for~\eqref{Riem} with $\gamma = 1/3$.}\\[1.0ex]
%In fact, here we are only interested in the projection of this restriction to the $(t, x, r, k)$-space, since such a projection is well defined and determine transformations of the rest internal coordinates (in terms of flow).}
\remarka{One can use suitable initial conditions to determine functions $g$ and $h$. However, additional invariants of characteristics are not defined
on some solutions. For example, solution to the problem~\eqref{initsol}(b) can not be reached using additional invariants on both characteristics (because $r_0'(1) = 0$).}

\subsection{Application of invariants in terms of a general solution}
%The second approach is based on using general solutions. It allows us to obtain exact solutions of some initial value problems for~\eqref{Eul_bar} in implicit form. 

There is another way to derive exact solutions for the system of equations~\eqref{Eul_bar} in cases where a general solution is known.
Below we demonstrate it for $\gamma = 5/3$.

The general solution has the form~\cite{Aksenov2} (in slightly different notations)
\begin{align}
\begin{aligned}
&t = \dfrac{108}{(k - r)^2}\Big(f_1' - f_2' + 2\,\dfrac{f_1 + f_2}{k - r}\Big),\\
&x = \dfrac{36}{(k - r)^2}\Big((2k + r)f_1' - (k + 2r)f_2' + 3(r + k)\,\dfrac{f_1 + f_2}{k - r}\Big).
\end{aligned}
\label{gensol}
\end{align}
Here $f_1 = f_1(r)$ and $f_2 = f_2(k)$ are arbitrary functions of their arguments. This general solution is written in the hodograph variables.
Nevertheless, using suitable initial conditions, we can define the functions $f_1$ and $f_2$. To this end, we substitute~\eqref{gensol} into the equation
\begin{align*}
x - kt + \dfrac{k - r}{2k_x} = h(k).
\end{align*}
Here $k_x$ is determined from~\eqref{gensol} using the implicit function theorem.
As a result, we obtain the following relation:
\begin{align*}
f_2''(k) = \dfrac{h(k)}{18}\,.
\end{align*}\\[1.0ex]
\remarka{Here the function $f_2(k)$ is determined from the function $h(k)$ up to an additive function of the form $C_1k + C_2$. Fortunately, these
constants $C_1$, $C_2$ are not essential for~\eqref{gensol} due to the transformation $f_1\mapsto f_1 - C_1r - C_2$.}\\[1.0ex]
Now the general solution~\eqref{gensol} allows to define $f_1(r)$ from initial conditions (if possible).

For instance, consider the initial conditions~\eqref{initsol}(a).
One can define
\begin{align*}
f_2(k) = \dfrac{k^3}{108} - \dfrac{k\arctan(k)}{36} + \dfrac{\ln(1 + k^2)}{72}\,.
\end{align*}
Substituting this into~\eqref{gensol} together with $t = 0$, $r = r_0(x)$, $k = k_0(x)$, we get the identity
\begin{align*}
\begin{aligned}
f_1\Big(\dfrac{x^3 + x + 1}{1 + x^2}\Big) = \ &
\dfrac{x^3 + x + 1}{36(1 + x^2)}\, \Big(\arctan(x) - \dfrac{x}{1 + x^2}\Big) -
\dfrac{\ln(1 + x^2)}{72} - \dfrac{x^3}{108}\,.
\end{aligned}
\end{align*}
Replacing
$$
x = \dfrac{r + s_{+} + s_{-}}{3},\quad
s_{\pm} = \sqrt[3]{r^3 + 9r - \frac{27}{2} \pm \frac{3}{2}\sqrt{12(r^4 - r^3 + 2r^2 - 9r) + 93}}\,,
$$
we obtain the corresponding function $f_1(r)$ and the desired solution in implicit form.\\[1.0ex]
\remarka{Direct application of the general solution~\eqref{gensol} to a given Cauchy problem is difficult, since one has to solve the ODE system for functions $f_1$, $f_2$ of different arguments.}

%\subsection{Application of the characteristic flow}
%The third approach is based on using invariants of only one characteristic distribution. It allows us to reduce some initial value problems for~\eqref{Eul_bar} to systems of ordinary differential equations. This approach is less restrictive than the first and second approaches.

\section{Geometric interpretation}\label{Geom}
\subsection{On characteristic distributions}

It is convenient to consider differential equations together with all their differential consequences. Below we deal with the infinite prolongation of the system~\eqref{Eul_bar} and denote it by $\mathcal{E}$.
\begin{align*}
\mathcal{E}\colon\qquad
\begin{aligned}
&F_1 = 0,\quad D_x(F_1) = 0,\quad D_t(F_1) = 0,\quad D^2_x(F_1) = 0,\quad \ldots\\
&F_2 = 0,\quad D_x(F_2) = 0,\quad D_t(F_2) = 0,\quad D^2_x(F_2) = 0,\quad \ldots
\end{aligned}
%\label{gen}
\end{align*}
Here $F_1 = u_t + uu_x + c^2\rho^{-1}\rho_x$, $F_2 = \rho_t + u \rho_x + \rho u_x$.

Let us briefly recall a few facts from the geometry of jet spaces (see, e.g.~\cite{VinKr}).
Jet spaces $J^{k}(n, m)$ arise together with the projections 
\begin{align*}
\pi_{q + k,\, q}\colon J^{q + k}(n, m) \to J^{q}(n, m)\,,\qquad \pi_k\colon J^{k}(n, m) \to M\,,\qquad k = \infty,\, 0,\, 1,\, \ldots
\end{align*}
Here $M$ is the $n$-dimensional space of independent variables. Besides, there is a
natural way to represent a point $p_k\in J^{k}(n, m)$ by a pair of the form $(p_{k-1}, L_{p_k})$, where $p_{k-1} = \pi_{k,\, k-1}(p_k)$ and
\begin{align*}
L_{p_k}\subset T_{p_{k-1}} J^{k-1}(n, m)\,,\qquad
d\pi_{k-1}(L) = T_{\pi_{k}(p_k)}M\,,\qquad \dim L_{p_k} = n\,.
\end{align*}
Here $L_{p_k}$ is defined as the plane tangent to the graph of the $(k-1)$-jet of some section $s$ which represents $p_k$.
Furthermore, fibers of projections of the form $\pi_{k,\, k-1}$ are affine spaces.\\

Consider a system of differential equations at $J^k(n, m)$ and its infinite prolongation $\mathcal{S}\subset J^{\infty}(n, m)$.
Let $e$ be a point of $\mathcal{S}$, $\xi$ be a covector, $\xi\in T_x^{*} M$, $x = \pi_{\infty}(e)$. Denote by $e_{i}$ the corresponding projection $\pi_{\infty,\, i}(e)$. \\[0.5ex]
\definitiona{A nonzero covector $\pi_{\infty}^*(\xi)$ (at $e$) is characteristic for $\mathcal{S}$ if there exists a line
$$
l\ \subset\ \pi^{-1}_{k,\, k - 1}(e_{k - 1})
$$
tangent to the projection $\pi_{\infty,\, k}(\mathcal{S})$ at $e_k$ such that for each point $p_{k}\in l$ the following condition is satisfied:
\begin{align*}
d\pi_{k - 1}(L_{p_{k}}\cap L_{e_{k}}) = \{v\in T_x M\colon \langle \xi, v\rangle = 0\}.
\end{align*}
}\\[-2.5ex]
%Among other things, 
%Such characteristic covectors are related to non-uniqueness of lift of a smooth initial data to the system $\mathcal{S}$. 
%For the system~\eqref{Eul_bar} this definition reduces to the classical one.
\remarka{\label{Subs_char}
It is worth mentioning that each characteristic covector for a subsystem $\mathcal{N}\subset \mathcal{E}\ \, (\subset J^{\infty}(2, 2))$ is also characteristic covector for the system $\mathcal{E}$.}\\[1.0ex]
Characteristic covectors allow us to consider characteristic distributions.

%\begin{prop}
%Let $f_1$ and $f_2$ be functions on the system $\mathcal{E}$ such that their horizontal differentials $d_h f_1$ and $d_h f_2$ are linearly dependent. %Then $f_1$ and $f_2$ are invariants of a characteristic distribution of $\mathcal{E}$.
%\end{prop}
%\noindent
%\textbf{Proof.} First of all let us show that an additional constraint of the form $f_2 = g(f_2)$ does not lead to other nontrivial compatibility %conditions (for any function $g$).

\subsection{Geometric interpretation of the approach}
The characteristic distributions of $\mathcal{E}$ are spanned by the vector fields
\begin{align*}
X_{\pm} = \,\overline{\!D}_t + (u \pm c)\,\overline{\!D}_x\,.
%\label{Char_fie}
\end{align*}
Vector fields in total derivatives are tangential to any (infinitely prolonged) smooth solution of $\mathcal{E}$. If such a vector field has a flow, it moves an initial curve along the corresponding solution. Therefore, such vector fields are trivial symmetries of $\mathcal{E}$.

Suppose there is an $X_+$-invariant $I_+$ additional to the Riemann invariant $r$. Let us add an equation of the form $I_+ = g(r)$ to the system~\eqref{Eul_bar}. The relation $I_+ = g(r)$ leads to additional equations (differential consequences) on the manifold $\mathcal{E}$. Since the operator $r_x^{-1}\,\overline{\!D}_x$ acts on the set of $X_+$-invariants, the corresponding differential consequences (\textit{of different orders}) can be written in terms of invariants:
\begin{align}
I_{+} - g(r) = 0\,,\
\dfrac{1}{r_x}\,\overline{\!D}_x(I_+) - \dfrac{dg}{dr} = 0\,,\, \ldots\,,\
\Big(\dfrac{1}{r_x}\,\overline{\!D}_x\Big)^{s}(I_+) - \dfrac{d^sg}{dr^s} = 0\,,\, \ldots
\label{conseq}
\end{align}
Moreover, since the relation $\,\overline{\!D}_t(f) = -(u + c)\,\overline{\!D}_x(f)$ holds for any $X_+$-invariant $f$, there are no nontrivial consequences additional to those listed above. Then the corresponding overdetermined system of equations does not have nontrivial compatibility conditions.

The application of invariants to construction of smooth solutions without gradient catastrophe for~\eqref{Eul_bar} is based on the following observation. Geometry of the system~\eqref{Eul_bar} with the additional equation $I_+ = g(r)$ becomes similar to geometry of a scalar first-order equation. This overdetermined system also admits a characteristic (trivial) symmetry with the flow. According to the remark~\ref{Subs_char}, at each point this symmetry coincides with some characteristic vector for $\mathcal{E}$.
However, most of the $X_+$-invariants become functionally dependent due to 
differential consequences~\eqref{conseq} unlike $X_-$-invariants. So, the restriction of the trivial symmetry $X_-$ to the infinite prolongation of the overdetermined system under consideration has the flow. Its projection onto some finite jets also has the flow. This flow moves an initial curve along the corresponding solution.\\[1.0ex]
\remarka{Invariants of characteristics are related to conservation laws. Namely, if $f_1$ and $f_2$ are invariants of the same characteristic distribution of $\mathcal{E}$, the horizontal differential forms~\cite{VinKr} 
$d_h f_1$ and $d_h f_2$
are linearly dependent. Hence, for any functions $G_1$, $G_2$ of invariants of the same characteristic distribution of $\mathcal{E}$, the following identity holds:
$$
d_h \big(G_1\, d_h f_1 + G_2\, d_h f_2\big) = 0\,.
$$ 
Thus, the horizontal forms $G_1\, d_h f_1 + G_2\, d_h f_2$ represent conservation laws for $\mathcal{E}$.}

\section{Conclusion}
One-dimensional equations of isentropic gas motion are Darboux integrabile for some equations of state. Additional invariants of characteristics made it possible to reduce such systems to ordinary differential equations and construct solutions without gradient catastrophe. The results of the investigation can be used in approbation of numerical methods for solving the equations of gas dynamics.

\section*{Declaration of competing interest}

The authors declare that they have no known competing financial interests or personal relationships that could have appeared to influence the work reported in this paper.

\section*{Acknowledgement}

O.V. Kaptsov acknowledges the support from the Krasnoyarsk Mathematical Center and the funding by the Ministry of Science and Higher Education of the Russian Federation in the framework of the establishment and development of regional Centers for Mathematics Research and Education (Agreement No. 075-02-2022-873).

\end{document}